\documentclass[prd,aps,showpacs,preprintnumbers,amssymb]{revtex4}
\usepackage{graphicx}% Include figure files
\usepackage{epsf}
\usepackage{amsmath}
\usepackage{epstopdf}
\usepackage{multirow}
\usepackage{bm}
\usepackage{dcolumn}

\def\rpi{{\vec r}^{\,\prime}_i}
\def\vpi{{\vec v}_i}
\def\ppi{{\vec p}_i}

\def\rdpi{{\vec {\dot r}}^{\,\prime}_i}

\def\om{{\vec \omega}}
\def\omd{{\vec {\dot \omega}}}

\def\rp{{\vec r}^{\,\prime}}
\def\rdp{{\vec {\dot r}}^{\,\prime}}

\newcolumntype{M}[1]{>{\centering\arraybackslash}m{#1}}
\newcolumntype{N}{@{}m{0pt}@{}}

\begin{document}

\title{%
\hfill{\normalsize\vbox{%
\hbox{}
 }}\\

\vskip 1cm

{Non-inertial torques and the Euler equation}}

\author{Amir H. Fariborz
\footnote[1]{Email:
fariboa@sunypoly.edu}}

\affiliation{Department of Mathematics and Physics, SUNY Polytechnic Institute, Utica, NY 13502, USA}

\date{\today}

\begin{abstract}
	
The fundamental equation describing the rotational dynamics of a rigid body is ${\vec \tau}=d{\vec L} / dt$ which is a straightforward consequence of the Newton's second law of motion and is only valid in an inertial coordinate system.  While this equation is written down by an inertial observer, for practical purposes, it is worked out within a non-inertial ancillary coordinate system which is typically fixed in the rigid body.  This results in the famous Euler equation for rotation of the rigid bodies.   We show that it is also possible to describe the rotational dynamics of a rigid body from the point of view of a {\it non-inertial} observer (rotating with the ancillary coordinate system fixed in the rigid body), provided that the non-inertial torques are taken into account.  
We explicitly calculate the non-inertial torques and express them in terms of physical characteristics of the rigid body.  We show that the resulting dynamical equations exactly recover the Euler equation.

\end{abstract}
\maketitle

%\section{Introduction}

% Computer graphics: 21_MS_JS,13_JB_KE
% ======  Mathematical aspects: \cite{97_MA_FP}-\cite{21_JC_PM}
% ===============   broader studies:  \cite{97_ML}-\cite{02_AB_IM}

Dynamics of rigid bodies is well studied and presented in standard textbooks \cite{FC,G},  and despite being an old subject,   has resulted in a wide  range of modern investigations such as their mathematical underpinnings \cite{97_MA_FP}-\cite{21_JC_PM}, applications in computer graphics \cite{21_MS_JS}-\cite{13_JB_KE} and in various broader studies \cite{97_ML}-\cite{02_AB_IM}.  However, in this short article we have a rather narrow focus on the dynamical equation itself and how it can be derived from the non-inertial torques. 

In studying the rotation of rigid bodies, the typical approach is to employ appropriate rotating coordinate systems to formulate the rotational dynamics \cite{FC,G}.  This rotating system is only used as an ancillary system and the dynamical equations are in fact worked out from the point of view of an inertial observer.   Since the non-inertial ``forces'' are not relevant for the inertial observer, therefore,  in such approaches only the real physical forces are pertinent to the rotational dynamics.   \\

On the contrary,  if we want to study the  dynamics of  rotation from the point of view of a non-inertial observer, we cannot ignore the effects of the non-inertial torques.    For example, in a high precision  study of the motion of a spinning top in a coordinate system fixed on the surface of the Earth, the non-inertial torques become relevant (note that we typically ignore these  non-inertial torques only because the effect of the Earth rotation is small in this system, and not because they are equal to zero). If  a high precision experimental study of a spinning top (or similar systems) is conducted, the effects of the Earth rotation should be taken into account.\\

We study the rotational dynamics written by a non-inertial observer inside a rotating frame of reference. Evidently,  in such an inherently non-inertial coordinate system,  the non-inertial forces would have to be dealt with and consequently the translational dynamics of point particles inside the rigid body obeys a more complicated form of the Newton’s second Law of motion than the one that  we are used to in inertial systems.  Likewise, in working out the rotational dynamics (from the point of view of this non-inertial observer), the non-inertial torques become important.\\

To calculate the non-inertial torques on a rigid body, we can first  calculate the torques on the individual constituent particles inside the rigid body and then sum them up.  For an $i$-th particle inside a rigid body,  the Newton’s second law becomes (written by the non-inertial observer):

\begin{eqnarray}
{\vec F}_i + {\vec F'}_{Cor,i} + {\vec F'}_{trans,i}  + {\vec F'}_{centrif,i} = m_i\, {d^2{\rpi}\over dt^2}
\end{eqnarray}
where the first term on the left hand side is the sum of all physical forces, followed by the non-inertial ``forces'' of Coriolis, transverse and  centrifugal: 
\begin{eqnarray}
{\vec F'}_{Cor,i} &=& -2 \, m_i \, \om \times \rdpi\\
{\vec F'}_{trans,i} &=&  - m_i\, \omd \times \rpi \\
{\vec F'}_{centrif,i} &=&  - m_i\, \om \times \left( \om \times \rpi \right) 
\end{eqnarray}

To analyze the rotational dynamics of a rigid body in a rotating frame,  the torques produced by each of these forces should be taken into account.   These torques are simply calculated (within the rotating system fixed on the rigid body) as:
\begin{eqnarray}
{\vec \tau}_{Cor,i} &=& -2 \, m_i \,\rpi \times\left( \om \times \rdpi\right)\\
{\vec \tau}_{trans,i} &=&  - m_i\, \rpi \times \left(\omd \times \rpi \right)\\
{\vec \tau}_{centrif,i} &=&  - m_i\, \rpi \times \left[ \om \times \left( \om \times \rpi \right)\right] 
\end{eqnarray}
In order to calculate the net non-inertial torque,  it is convenient to rewrite these torques in a form that can be added up more effectively.  Invoking the Jacobi identity, we can rewrite the Coriolis torque as:
\begin{eqnarray}
{\vec \tau}_{Cor,i} &=& -2 \, m_i \,\rpi \times\left( \om \times \rdpi\right)\nonumber \\
&=& - \, m_i \,\rpi \times\left( \om \times \rdpi\right) - \, m_i \,\rpi \times\left( \om \times \rdpi\right)\nonumber \\
&=& - \, m_i \,\rpi \times\left( \om \times \rdpi\right) + m_i \,\om \times \left(\rdpi \times \rpi \right) + m_i\, \rdpi \times \left(\rpi \times \om\right)\nonumber \\
&=& - \, m_i \,\rpi \times\left( \om \times \rdpi\right) - m_i \,\om \times \left(\rpi \times \rdpi \right) - m_i\, \rdpi \times \left(\om \times \rpi\right)
\end{eqnarray}
The centrifugal torque is a quadruple cross product and can be rearranged (using Jacobi identity again) as:
\begin{eqnarray}
{\vec \tau}_{centrif,i} &=& - m_i\, \rpi \times \left[ \om \times \left( \om \times \rpi \right)\right] \nonumber \\
&=&
m_i\, \om \times \left[ \left(\om \times \rpi \right) \times \rpi \right] +
m_i\, \left(\om \times\rpi \right) \times \left(\rpi \times \om \right)   \nonumber \\
&=& m_i\, \om \times \left[ \left(\om \times \rpi \right) \times \rpi \right]
\end{eqnarray}
This gives:
\begin{eqnarray}
{\vec \tau}_{non-inertial,i} &=& {\vec \tau}_{Cor,i} + {\vec\tau}_{trans,i}  + {\vec\tau}_{cent,i} 
\nonumber \\
 &=& 
 - \, m_i \,\rpi \times\left( \om \times \rdpi\right) - m_i \,\om \times \left(\rpi \times \rdpi \right) - m_i\, \rdpi \times \left(\om \times \rpi\right)
 - m_i\, \rpi \times \left(\omd \times \rpi \right) 
\nonumber \\
&&
+
m_i\, \om \times \left[ \left(\om \times \rpi \right) \times \rpi \right]
\nonumber \\
&=&
-m_i \left[ \rdpi \times \left(\om \times \rpi\right) + \rpi \times \left(\omd \times \rpi \right) +
 \rpi \times\left( \om \times \rdpi\right) \right] 
 -
 m_i\, \om\times \left[ \rpi \times \rdpi - \left(\om \times \rpi\right) \times\rpi \right]
\nonumber \\
&=& -m_i \, 
\left\{ {d\over dt} \left[\rpi\times \left(\om\times\rpi\right) \right]
+ \om\times \left[ \rpi \times \left(\om \times \rpi\right) \right] \right\} - m_i \om \times \left( \rpi\times\rdpi\right)
\end{eqnarray}
Adding and subtracting time derivative of $\rpi\times\rdpi$ will result in:
\begin{eqnarray}
{\vec \tau}_{non-inertial,i} &=& -m_i \, 
\left\{ 
       {d\over dt} \left[ \rpi\times \left(\rdpi + \om\times\rpi\right) 
                   \right]
       + \om\times  \left[ 
                          \rpi \times \left(\rdpi + \om \times \rpi\right) 
                   \right]
                   \right\} 
+ m_i {d\over dt} \left( \rpi\times\rdpi\right)
\end{eqnarray}
The terms in the parentheses  yield the total velocity of the particle $i$ with respect to the inertial frame:
\begin{eqnarray}
{\vec \tau}_{non-inertial,i} &=& -m_i \, 
\left[ 
        {d\over dt} \left( \rpi\times \vpi \right)
        + \om\times \left( \rpi\times \vpi \right) 
\right] 
+ m_i {d\over dt} \left( \rpi\times\rdpi\right)
\nonumber \\
 &=& - \, 
\left[ 
        {d\over dt} \left( \rpi\times \ppi \right) 
        + \om\times \left( \rpi\times \ppi \right) 
\right]
+ m_i {d\over dt} \left( \rpi\times\rdpi\right)
\label{tau_NI}
\end{eqnarray}
The terms in the parentheses  are the  angular momentum of particle $i$ with respect to the inertial frame and therefore the terms inside the  square brackets are the time variation of angular momentum with respect to the inertial frame, i.e. 
\begin{eqnarray}
{d\over dt} \left( \rpi\times \ppi \right)
+ \om\times \left( \rpi\times \ppi \right) 
=
\left({{d{\vec L}_i}\over dt}\right)_{rot} + \om\times {\vec L}_i
=\left( {d{\vec L}_i\over dt}\right)_{inertial} = {\vec\tau}_{phys,i}
\label{dL_dL_rot}
\end{eqnarray}
The last term in (\ref{tau_NI}) is:
\begin{eqnarray}
m_i\, {d\over dt} \left( \rpi\times \rdpi \right)
= m_i \, {d\over dt} \rpi \times \rdpi +   m_i \, \rpi \, \times {d\over dt} \rdpi = \rpi \, \times 
m_i {d\over dt}\rdpi = \left( {d{\vec L'}_i\over dt}\right)_{rot}
\label{dLp_dt}
\end{eqnarray}

Substituting (\ref{dL_dL_rot}) and (\ref{dLp_dt}) into (\ref{tau_NI}), we get:
\begin{eqnarray}
\left( {d{\vec L'}_i\over dt}\right)_{rot} = {\vec \tau}_{non-inertial,i} +
 {\vec \tau}_{phys, i}
\end{eqnarray}

The subscript $i$ in the above equation runs over all points inside the rigid body. Summing  over the dynamical equations written for every such point inside the rigid body we get:

\begin{eqnarray}
\sum_i \left( {d{\vec L'}_i\over dt}\right)_{rot} = \sum_i {\vec \tau}_{non-inertial,i} + \sum_i
 {\vec \tau}_{phys, i}
\end{eqnarray}

or:

\begin{eqnarray}
 \left({{d{\vec L'}}\over dt}\right)_{rot} = {\vec \tau}_{non-inertial} + {\vec \tau}_{phys}
\label{Rot_Eq}
\end{eqnarray}

Therefore,  to study the rotational dynamics of a rigid body from the point of view of the observer in a rotating frame, the non-inertial torques are relevant and have to be taken into account.   However,  when the rotational dynamics is studied in an inertial frame using a non-inertial frame just as an ancillary tool,  the non-inertial torques do not appear in the dynamical equations. In the inertial coordinate system, the rotational dynamics is given by Euler's equation:
\begin{eqnarray}
{\vec \tau}_{phys} =
\left( 
{{d{\vec L}}\over dt}  
\right)_{rot} + {\vec \omega'} \times {\vec L}
\label{E_Eq}
\end{eqnarray}

Rewriting (\ref{Rot_Eq}) in the form
\begin{eqnarray}
{\vec \tau}_{phys} = 
\left({{d{\vec L'}}\over dt}\right)_{rot} - {\vec \tau}_{non-inertial} 
\label{Rot_Eq2}
\end{eqnarray}

A comparison of equation (\ref{E_Eq}) with (\ref{Rot_Eq2}) shows that the right hand sides of these two equations should be identical.   We show that this is indeed the case. 

The non-inertial torques on the entire rigid body can be computed through the following integrations in the continuous limit  
\begin{eqnarray}
{\vec \tau}_{Cor} &=& -2 \, \int \rp \times\left( \om \times \rdp\right)\, dm'
\\
{\vec \tau}_{trans} &=&  - \int \rp \times \left(\omd \times \rp \right)\, dm'
\\
{\vec \tau}_{cent} &=&  -\int \rp \times \left[ \om \times \left( \om \times \rp \right)\right]\, dm' 
\end{eqnarray}
Note that $\omega$ is not an intrinsic characteristic of the rigid body, therefore, it is useful to factor it out.  We can conveniently achieve this in component form.   The $k$-th component of the  Coriolis torque can be written as:
\begin{eqnarray}
\tau_{Cor,k} &=&
-2\,\sum_{m=1}^3\sum_{j=1}^3\sum_{l=1}^3\sum_{n=1}^3  \int \epsilon_{kmj}\, x'_m\, \epsilon_{jln}\, \omega_l\, {\dot x'}_n\, dm' \nonumber \\
&=& -2\, \sum_{m=1}^3\sum_{j=1}^3\sum_{l=1}^3\sum_{n=1}^3 \epsilon_{kmj}\, \epsilon_{jln}\, \omega_l\, \int x'_m \,  {\dot x'}_n\, dm'
\end{eqnarray}
where $\epsilon$ is the Levi-Civita tensor.  Note that, for more transparency,  we have explicitly  displayed the summations (using Einstein summation notation simplifies these formulas). 
Defining a new tensor:
\begin{equation}
J_{mn} = \int x'_m \,  {\dot x'}_n\, dm',
\end{equation}
which measures the ``distribution of velocity'' inside the rigid body,  we can write:
\begin{eqnarray}
\tau_{Cor,k} &=& -2\,\sum_{m=1}^3\sum_{j=1}^3\sum_{l=1}^3\sum_{n=1}^3 \epsilon_{kmj}\, \epsilon_{jln}\, \omega_l\, J_{mn}.
\end{eqnarray}
Using the identity:
\begin{equation}
\sum_{j=1}^3\epsilon_{kmj}\, \epsilon_{jln} = \delta_{kl} \delta_{mn} -  \delta_{kn} \delta_{lm},
\end{equation}
we can finally  write
\begin{equation}
\tau_{Cor,k} = -2\, \left( \omega_k\, 
\sum_{m=1}^3 J_{mm} - \sum_{m=1}^3\omega_m J_{mk} \right).
\label{Coriolis}
\end{equation}

%%%%%%%%%%%%%%%%%%%%%%%%%%%%%%%%%%%%%%%%%%%%%%%%%%%%%%%%%%%%%%%%%%%%%%%%%%%

Similarly, we can work out the $k$-th component of transverse torque: 

\begin{eqnarray}
\tau_{trans,k} &=& -\sum_{m=1}^3 \,\sum_{j=1}^3 \, \sum_{l=1}^3 \, \sum_{n=1}^3 
\epsilon_{kmj} \, \epsilon_{jln} \, 
{\dot \omega}_l\,\int x'_m\, x'_n\, dm' \nonumber \\
&=& -\sum_{m=1}^3 \, \sum_{l=1}^3 \, \sum_{n=1}^3 
\left(\delta_{kl}\delta_{mn} - \delta_{kn}\delta_{ml}\right) \, 
{\dot \omega}_l\,\int x'_m\, x'_n\, dm'.
\end{eqnarray}
The integrals are related to the products of inertia 
\begin{equation}
I_{mn} = - \int x'_m\, x'_n\, dm'.
\end{equation}
Therefore
\begin{eqnarray}
\tau_{trans,k} 
&=& \sum_{m=1}^3 \, \sum_{l=1}^3 \, \sum_{n=1}^3 
\left(\delta_{kl}\delta_{mn} - \delta_{kn}\delta_{ml}\right) \, 
{\dot \omega}_l\,I_{mn} \nonumber \\
&=& \sum_{m=1}^3 \, \left({\dot \omega}_k\,I_{mm} - {\dot \omega}_m\,I_{mk} \right).
\label{transverse}
\end{eqnarray}

The centrifugal torque is slightly more complicated because of the quadruple product.  Its $k$-th component is:   
\begin{eqnarray}
\tau_{Cent,k} &=& -\sum_{m=1}^3 \,\sum_{j=1}^3 \, \sum_{l=1}^3 \, \sum_{p=1}^3 \, 
\sum_{s=1}^3 \, \sum_{n=1}^3 \,
\epsilon_{kmj} \, \epsilon_{jlp} \,  \epsilon_{psn} \,
\omega_l\,\omega_s\, \int x'_m\, x'_n\, dm' 
\nonumber \\
&=& \sum_{m=1}^3 \,\sum_{j=1}^3 \, \sum_{l=1}^3 \, \sum_{p=1}^3 \, 
\sum_{s=1}^3 \, \sum_{n=1}^3 \,
\epsilon_{kmj} \, \epsilon_{jlp} \,  \epsilon_{psn} \,
\omega_l\,\omega_s\, I_{mn}
\nonumber \\
&=& \sum_{m=1}^3 \,\sum_{j=1}^3 \, \sum_{l=1}^3 \,  
\sum_{s=1}^3 \, \sum_{n=1}^3 \,
\epsilon_{kmj} \, \left(\delta_{js}\delta_{ln}-\delta_{jn}\delta_{ls}\right)
\omega_l\,\omega_s\, I_{mn}
\nonumber \\
&=& \sum_{m=1}^3 \,\sum_{j=1}^3  \, \sum_{n=1}^3 \,
\epsilon_{kmj} \, \left(\omega_j\,\omega_n - \omega^2 \delta_{jn}  \right)
\, I_{mn}
\nonumber \\
&=& \sum_{m=1}^3 \,\sum_{j=1}^3  \, \sum_{n=1}^3 \,
\epsilon_{kmj} \, \omega_j\,\omega_n\, I_{mn}
\label{centrifugal}
\end{eqnarray}

%\section{}

Using equations (\ref{Coriolis}), (\ref{transverse}) and (\ref{centrifugal}), we can now rewrite equation (\ref{dL_dL_rot}) within a coordinate system fixed in the rigid body.  Since in this system the rigid body is at rest, then:
\begin{eqnarray}
{d{\vec L}'\over dt} & = & 0\\
{\vec \tau}_{\rm Cor} & = & 0
\label{tau_Cor_abc}
\end{eqnarray}
In this system, the transverse torque becomes:
\begin{eqnarray}
\tau_{trans,k} &=& \sum_{m=1}^3 \left( {\dot \omega}_k\, I_{mm} - {\dot \omega}_m I_{mk}\right) \nonumber \\
&=&\left( \sum_{m=1}^3  {\dot \omega}_k\, I_{mm} \right) - {\dot \omega}_k I_{kk}\nonumber \\
&=& \sum_{m \ne k}^3  {\dot \omega}_k\, I_{mm} \nonumber \\ 
&=& - {\dot \omega}_k\, I_{k}
\end{eqnarray}

The Centrifugal torque in this system takes the form:
\begin{eqnarray}
\tau_{Cent,k} &=&
\sum_{m=1}^3 \,\sum_{j=1}^3  \, \sum_{n=1}^3 \,
\epsilon_{kmj} \, \omega_j\,\omega_n\, \delta_{mn}\, I_{mm} \nonumber \\
&=& 
\sum_{m=1}^3 \,\sum_{j=1}^3  \,
\epsilon_{kmj} \, \omega_j\,\omega_m\, I_{mm}
\label{tau_trans_abc}
\end{eqnarray}
It is straightforward to examine different components of the above double sum.  Let us work out $k=1$ component:
\begin{eqnarray}
\tau_{Cent,1} &=&
\sum_{m=1}^3 \,\sum_{j=1}^3  \,
\epsilon_{1mj} \, \omega_j\,\omega_m\, I_{mm}\nonumber \\
&=& 
\epsilon_{123} \, \omega_3\,\omega_2\, I_{22} + \epsilon_{132} \, \omega_2\,\omega_3\, I_{33} 
\nonumber \\
&=&
\omega_2\, \omega_3 \left( I_{22} - I_{33}\right)
\nonumber \\
&=&
\omega_2\, \omega_3 \left( I_{22} +I_{11} - I_{33}- I_{11} \right)
\nonumber \\
&=&
-\omega_2\, \omega_3 \left(I_3 - I_2 \right)
\label{tau_cent_abc}
\end{eqnarray}
in which $I_2=-(I_{11}+I_{33})$ is the moment of inertia about the second axis fixed in the rigid body (similarly for $I_3$).    
Likewise, the other components can be computed.  Now substituting Eqs.  (\ref{tau_Cor_abc}) to (\ref{tau_cent_abc}) into  (\ref{Rot_Eq2}) yields:
\begin{equation}
 \tau_{phys, 1} = {\dot \omega}_1\,I_1 + \omega_2\, \omega_3(I_3 - I_2)
\end{equation}
Similarly, we can calculate other components of the physical torque and arrive at:
\begin{equation}
\left[
\begin{array}{cc}
\tau_{phys, 1}\\
\tau_{phys, 2}\\
\tau_{phys, 3}
\end{array}
\right] =
\left[
\begin{array}{cc}
I_1 \, {\dot \omega}_1\\
I_2\, {\dot \omega}_2\\
I_3\, {\dot \omega}_3
\end{array}
\right] 
+
\left[
\begin{array}{cc}
\omega_2\, \omega_3  \left(I_3 - I_2\right)\\
\omega_3\, \omega_1  \left(I_1 - I_3\right)\\
\omega_1\, \omega_2  \left(I_2 - I_1\right)
\end{array}
\right] 
\end{equation}
which is exactly the Euler equation for the dynamics of  a rotating rigid body witten by an inertial observer  (using the system fixed in the rigid body as an ancillary coordinate system).  

In this work, we studies the rotational dynamics of a rigid body from the point of view of a non-inertial observer.  We showed that the non-inertial torques have to be taken into account in the dynamical equation of rotation written by the non-inertial observer. We explicitly calculated the non-inertial torques and expressed them in tensor form that factor out the angular velocity from the characteristics of the rigid body.   We then showed that the derived rotational dynamics, when applied within a coordinate system fixed in the rigid body,  exactly recovers the well-known Euler equation for the rotation of a rigid body and that the Coriolis torque does not play a role in this case.   It is interesting to investigate the Coriolis torque in more detail and examine cases where it may have nontrivial effects.

\end{document}